\documentclass[prl,twocolumn,floatfix,superscriptaddress,showpacs]{revtex4}

\usepackage{graphicx}

\newcommand{\be}{\begin{eqnarray}}
\newcommand{\ee}{\end{eqnarray}}
\def\ket#1{|#1\rangle}
\def\bra#1{\langle #1 |}  

\begin{document}

\title{Bloch oscillations in Fermi gases}

\author{M.\ Rodr\'\i guez}


\affiliation{Laboratory of Computational Engineering, P.O.Box 9203, 
FIN-02015 Helsinki University of Technology, Finland}

\author{P.\ T\"orm\"a}


\affiliation{Department of Physics, University of Jyv\"askyl\"a, P.O.Box 
35, FIN-40014 Jyv\"askyl\"a, Finland}

\date{\today}

\begin{abstract}

The possibility of Bloch oscillations for a degenerate and 
superfluid Fermi gas of atoms in an optical lattice is considered. 
For a one-component degenerate gas the oscillations are suppressed for high temperatures and band fillings. 
For a two-component gas, Landau criterion is used for specifying the regime where Bloch oscillations of the 
superfluid may be observed. We show how the amplitude of Bloch oscillations varies along the BCS-BEC crossover.       

\end{abstract}

\pacs{03.75.Ss, 03.75.Lm, 05.30.Fk, 32.80.-t, 74.25.-q}

\maketitle

The experimental realization of optical lattices for bosonic atoms
has led to several landmark experiments \cite{optlattices,bose0,bose1}. 
Very recently similar potentials have become available for
trapping the {\it fermionic} isotopes as well \cite{inguscio,rudi}. An 
increase in the superfluid transition temperature when using potentials 
created by standing light waves has been predicted \cite{Zoller}. 
For trapped cold atoms, the famous BCS-BEC crossover problem \cite{legget,nozi}
could be studied by tuning the interaction strength between the atoms using 
Feshbach resonances \cite{expfermi,rudi}. In optical lattices the whole 
BCS-BEC crossover could be scanned experimentally also in an even simpler way
by modulating the light intensity. 
We consider Bloch oscillations in these systems and show that they can be used as a tool for studying the crossover.  

Bloch oscillations are a pure quantum phenomenon occuring in a periodic potential. They have never been observed in a natural
lattice for electrons  as predicted in \cite{bo} because the scattering time of the electrons by 
lattice defects or impurities is much shorter than the Bloch period. 
However, Bloch oscillations have recently been observed in 
semiconductor superlattices 
\cite{sl}, 
for 
quasiparticles penetrating the cores of a vortex lattice in
a cuprate superconductor \cite{vortexbo}, and for
periodic optical systems such as waveguide arrays \cite{waveguide}. 
Also cold bosonic atoms and superfluids in optical lattices have been shown to be 
clean and controllable systems well suited for the observation of Bloch oscillations \cite{bose0,bose1}. 

Several novel aspects of the physics of Bloch oscillations arise for  
fermionic atoms in optical lattices. i) Impurity scattering can be made  negligible, and the particle
number controlled at will to produce any band filling. Even when Bloch oscillations were originally proposed 
for fermions, the effect of the Fermi sea has not played a major role. Due to impurity and defect scattering, the studies of
transport in presence of a constant force have focused on drift velocities rather than 
oscillations. In this letter we generalize the semiclassical single particle description of Bloch oscillations
to arbitrary band fillings. ii) The possibility of an oscillating fermionic 
superfluid becomes relevant. We derive the Landau criterion for the optical lattice imposing the Cooper 
pair size to be of the order or smaller than the lattice spacing. For solid state systems, the Cooper pair radius is usually 
much larger than the lattice spacing and periodicity 
irrelevant for the superfluid, therefore the system is treated as homogeneous 
when calculating supercurrents. We calculate the 
superfluid velocity in the {\it periodic} potential and show that pairing, leading to smoothening of the Fermi edge, suppresses 
Bloch oscillations.

Using six counter-propagating laser beams of wavelength $\lambda$, 
an isotropic 3D simple cubic lattice potential can be created which is of the form
\begin{equation}
V({\bf r}) = V_0 \left[\cos^2\left(\frac{\pi x}{a}\right) +  \cos^2\left(\frac{\pi y}{a}\right) + 
\cos^2\left(\frac{\pi z}{a}\right)\right] , \label{laserpot}
\end{equation}
where $V_0$ is proportional to the laser intensity and $a=\lambda/2$.
With the Bloch ansatz the Schr\"odinger equation leads to a band structure in the energy spectrum $\varepsilon_n({\bf k})$.
One-component degenerate Fermi gas at low temperatures can be considered as 
{\it non-interacting} since p-wave scattering is negligible and s-wave scattering suppressed by Fermi statistics. 
We are interested in high enough values of $V_0$  
such that tunneling is small and 
tight binding approximation can be applied. The dispersion relation for the lowest band becomes
$\varepsilon({\bf k})=J[3-\cos(k_xa)-\cos(k_ya)-\cos(k_za)]$, where
the band width $J= \frac{2}{\sqrt{\pi}}E_R \left(
\frac{V_0}{E_R}\right)^{3/4}\exp\left(-2\sqrt{\frac{V_0}{E_R}}\right)$ is obtained using the WKB-approximation 
and $E_R$ is the recoil energy of the lattice \cite{Zoller}.

\begin{figure}
\includegraphics[scale=0.4]{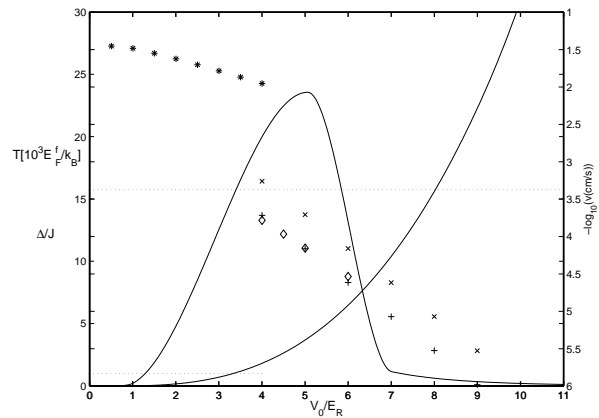}
\caption{The transition temperature, Landau criterion at T=0 and the amplitude 
of the velocity Bloch oscillations for $^6Li$ atoms in hyperfine states 
with scattering length $a_s=-2.5\cdot10^3a_0$ for a half filled 3D $CO_2$ laser
lattice ($a=10^5a_0$)as a 
function of the lattice depth. The amplitudes of the oscillations at 
$T= 2/3T_c^{max}$  (horizontal line) are denoted by $\ast$ for the 
normal state oscillations, 
$\diamond$ for the superfluid velocity at T=0 Eq.(\ref{vs}) and in 
the boson limit Eq.(\ref{blim}) pair size $l=a/3$ by $\times$ and 
$+$ for pair size $l=a/4$. The Landau criterion condition
Eq.(\ref{vcritical3}) requires $\Delta/J>1$ for the half filled band.
Here $E_R$ is the recoil energy and $E_F^f$ is the Fermi energy for free fermions with the same density.}
\label{landaucriteria}

\end{figure}

In a two-component Fermi gas, atoms in two different hyperfine states ($"\downarrow,\uparrow"$) may interact with each other. 
The interaction can be assumed pointlike, characterized by a scattering length $a_S$. The system Hamiltonian
$
\hat{H}=\sum_\alpha \int d^3{\bf r}\hat{\psi}_\alpha^\dagger({\bf r})(T+V)\hat{\psi}_\alpha({\bf r})-|g|\int d^3{\bf r}\hat{\psi}_\uparrow^\dagger\hat{\psi}_\downarrow^\dagger\hat{\psi}_\downarrow\hat{\psi}_\uparrow$, where $g = 4\pi
\hbar^2 a_S/m$ 
can then be mapped to the attractive Hubbard model
$
\hat{H}=J\sum_{\langle i,j \rangle \sigma}\hat{c}^\dagger_{i\sigma}\hat{c}_{j\sigma}-
U\sum_j \hat{c}^\dagger_{j\uparrow}\hat{c}^\dagger_{j\downarrow}
\hat{c}_{j\downarrow}\hat{c}_{j\uparrow}$,
where $U= E_R \frac{|a_s|}{a}\sqrt{8\pi}\left( \frac{V_0}{E_R}\right)^{3/4}$. 
Note that  
the BCS ($J>> U$) to BEC ($U>>J$) cross-over can be controlled by $V_0$ alone. One-band description is used in the Hubbard model 
also in the case of strong interactions \cite{long}. 
We define the limits of the one-band approximation for the physical potential Eq.(\ref{laserpot}) by demanding
the lowest band gap to be bigger than the effective interaction $U$ (note that $U>|g|$ for the parameters of interest). The band gap can be  
estimated by approximating the cosine potential well by a quadratic one. Demanding the corresponding harmonic
oscillator energy to be greater than $U$ gives 
the condition 
$\frac{V_0}{E_R} < \frac{1}{4 \pi^2}\left(\frac{a}{|a_s|}\right)^4$. Since $a>|a_S|$ imposed by considering on-site interactions only, the condition is easily valid in general, and for the parameters of Fig.\ \ref{landaucriteria} in particular. Estimates made using exact numerical band gaps in 1D support this argument. One-band approximation is sufficient because larger $V_0$ means steeper optical potential wells which not only increase the effective interaction $U$ but also the band gaps.

Bloch oscillations for a single atom can be characterized considering
the mean velocity of a particle in a Bloch state 
${\bf v}(n,{\bf k})=\bra{n,{\bf k}}\dot{{\bf r}}\ket{n,{\bf k}}$
given by 
\begin{equation}
{\bf v}(n,{\bf k})=\frac{1}{\hbar}\nabla_{\bf k} \varepsilon_n({\bf k}) . 
\label{bv}
\end{equation}  
When a particle in the Bloch state 
$\ket{n,{\bf k}_0}$ is adiabatically affected by a constant external 
force ${\bf F}=F_x {\bf \hat{x}}$ weak enough not to induce interband transitions, it 
evolves up to a phase factor into the state $\ket{n,{\bf k}(t)}$ according to
${\bf k}(t)={\bf k}_0+\frac{{\bf F}ta}{\hbar}$. 
The time evolution has a period 
$\tau_B=h/(|F_x|a)$, corresponding to 
the time required for the quasimomentum to scan the whole Brillouin zone. 
If the force is applied adiabatically, it provides momentum to the system but not energy because 
the effective mass (given by $m(\varepsilon)^{-1}= 
\frac{1}{\hbar^2}\frac
{\partial^2 \varepsilon}{\partial k^2 }$) is not always positive. 
For optical lattices the force (or tilt: $V= {\bf F \cdot r}$ term in the Hamiltonian) 
can be realized by accelerating the lattice \cite{bose0,bose1}. 
Using the tight-binding dispersion 
relation the velocity of an atom oscillates like
\begin{equation}
 v_{x}(t)=\frac{Ja}{\hbar}\sin(k_{0x}a+F_xta/\hbar).  \label{onemode}
\end{equation}

\begin{figure}
\includegraphics[scale=0.33]{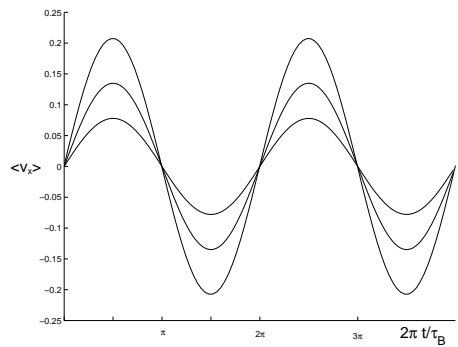}
\caption{The average velocity in $\frac{Ja}{\hbar}$ units as a function of
 time for a half filled band. The plotted lines correspond to atoms in
 the normal state at different temperatures $T=J$, $0.5J$ and 
$ \le 0.1J$. Bigger amplitudes correspond to lower temperatures.}
\label{normalygamma}
\end{figure}

For cold bosonic atoms and condensates \cite{bose0,bose1} nearly all of the population is in the 
lowest mode of the optical potential, Eq.(\ref{onemode}) therefore describes the oscillation of the whole gas.
We generalize the result for
the case when many momentum states of the band (at T=0, 
the states with wave vector ${\bf k} \le {\bf k}_F$) are occupied.
We calculate the velocity of the whole gas as the average over
the normalized temperature-dependent distribution function (the Fermi distribution $f$) of 
the particles:	
\begin{equation}
\langle {v_x}(t) \rangle=\frac{1}{\hbar}\sum_{{\bf k}_0} f({\bf k_0})
\nabla_{k_{0x}} \varepsilon({\bf k}_0+{\bf F}t/\hbar). \label{1cT}
\end{equation}
Using the tight-binding dispersion relation for the Bloch energies we 
obtain the oscillations shown in Fig.\ \ref{normalygamma}. 
At T=0, Eq.(\ref{1cT}) reduces to
\begin{equation}
\langle v_x (t) \rangle = 
\frac{Ja}{\hbar}\frac{\sin(k_{xF}a)}{k_{xF}a}\sin\left(\frac{F_xta}{\hbar}
\right).
\end{equation}  
This shows that macroscopic coherent oscillation 
effect can still be observed if the band is not full, but the amplitude is 
suppressed by the band filling $k_{xF}a$. The effect of the temperature 
can be seen in Fig.\ 
\ref{normalygamma}: the amplitude starts to decrease 
at temperatures of the order $T \geq 0.1J$
but is still non-negligible at half J. 
These results are valid for the one-component 
degenerate Fermi gas at low temperatures. In the two-component Fermi gas, 
atoms in the different hyperfine states interact with each other which
may lead to a superfluid state. Above $T_c$, weak
interactions can be described by a mean field shift in the chemical potential, leading to
no qualitative changes in Bloch oscillations. Inelastic scattering and consequent damping of Bloch oscillations
can be described e.g. by balance equations \cite{rusos}. In the following we consider the 
superfluid case where qualitative changes are expected.

In order to observe Bloch oscillations of the superfluid Fermi gas, the critical velocity of the 
superfluid should not be reached before the edge of the Brillouin zone. A BCS-superconductor can carry a 
persistent current $q$ until a critical velocity, $v_c=\frac{\Delta}{p_F}$.
For higher current values, even at $T=0$, it might be 
energetically favorable to break Cooper pairs and create a pair of
quasiparticles \cite{perscur}.  
This costs $2\Delta$ in binding energy and decreases the Bloch energy by 
$|\xi_{k+q}-\xi_{k-q}|\equiv 2|E_D|$. Therefore, for the current to be stable
$|E_D| < \Delta$. 
This is the Landau criterion of superfluidity.
For a tight binding lattice dispersion relation, we 
rewrite the condition 
as $J\sin(qa)\sin{k_Fa} < {\Delta}$. 
To complete a Bloch oscillation, $\sin(qa)$ should achieve its maximum value $1$, i.e.  
\begin{equation} 
\sin{k_Fa} <\Delta/J. \label{vcritical3}
\end{equation}
For weak coupling, $\Delta/J$ is given by the BCS theory, and 
in the attractive Hubbard model in the strong coupling limit the gap at 
$T=0$ is given by 
$\Delta=\frac{U}{2}$ for half filling \cite{long}. Using these estimates, we show
in Fig.\ \ref{landaucriteria} 
the relation (\ref{vcritical3}) for a gas of $^6Li$ atoms together with the transition temperature. 
To relate the criterion to the Cooper pair size, 
we rewrite Eq.(\ref{vcritical3}) in terms 
of the BCS coherence length $\xi_0=\frac{\hbar v_F}{\pi \Delta}$ and insert
$J\sin(k_Fa)= \hbar v_F/a$ which yields 
$ \xi_0 < a/\pi$.
The observation of Bloch oscillations is thus restricted to superfluids
with BCS coherence length smaller than the lattice periodicity. 
This is the intermediate -- strong coupling regime. 
The length argument can be also
understood by thinking that the pairs have to be smaller than the lattice sites
in order to see it as a periodic potential. 

For calculating the superfluid velocity a space dependent description of the
superfluid has to be used. We combine the BCS ansatz with the Bloch ansatz for the lattice potential
using the Bogoliubov -- de Gennes (BdG) equations \cite{BdG}.
As given by the Landau criterion above, the interesting regime is the 
intermediate -- strong coupling one. Note that even in the strong-coupling limit, the algebra of the BCS theory 
can be applied to all coupling strengths \cite{nozi,BCSalgebra} together with an extra definition for 
the chemical potential which in the weak 
coupling limit is given just by the Fermi energy of the non-interacting gas.
The BdG equations are:
\be
\left(\begin{array}{cc} H({\bf r})-\mu  & \Delta({\bf r})\\
\Delta({\bf r})^\ast & -[H({\bf r})-\mu] \end{array}\right)
\left(\begin{array}{c} u({\bf r})\\ v({\bf r}) \end{array}\right)=E
\left(\begin{array}{c} u({\bf r})\\ v({\bf r}) \end{array}\right) . 
\label{bdg}
\ee
When the external potential is periodic one can use the Bloch ansatz for 
$u$ and $v$ because by self-consistency the Hartree and pairing fields are 
also periodic. We obtain
\be
u_{\bf k}({\bf r})=e^{i{\bf k \cdot r}}\tilde{u}_{\bf k}\phi_{\bf k}({\bf r}) \quad ; \quad 
v_{\bf k}({\bf r})=e^{i{\bf k \cdot r}}\tilde{v}_{\bf k}\phi_{\bf k}({\bf r}) \label{uv1} \\
\Delta({\bf r})=\sum_{\bf k}|g|[1-2f(E_{\bf k})]u_{\bf k}({\bf r})v^\ast_
{\bf k}({\bf r}) , \label{gap1}
\ee
where $\phi_{\bf k}$ are the 
Bloch enveloping functions, such that $[H({\bf r})-\mu]\phi_{\bf k} 
e^{i{\bf k \cdot r}}=\xi_{\bf k}\phi_{\bf k} e^{i{\bf k \cdot r}}$.

To describe Bloch oscillations we impose the adiabatic condition, that is, momenta evolve according to $ {\bf k} \longrightarrow {\bf k} + {\bf F}ta/\hbar \equiv {\bf k} + {\bf q}$, i.e.\ we consider BCS state with a drift (again only in x-direction). The solutions of the BdG equations 
take the form 
\be
u^{\bf q}_{\bf k}({\bf r})&=&e^{i{\bf k \cdot r}}e^{i{\bf q \cdot r}}\tilde{u}^{\bf q}_{\bf k}\phi_{\bf k+q}({\bf r}) 
; 
v^{\bf q}_{\bf k}({\bf r})=e^{i{\bf k \cdot r}}e^{-i{\bf q \cdot r}}\tilde{v}^{\bf q}_{\bf k}\phi_{\bf k-q}({\bf r}) 
\nonumber \\
\Delta^{\bf q}({\bf r})&=&\sum_{\bf k}|g|[1-2f(E^{+{\bf q}}_{\bf k})]u^{\bf q}_{\bf k}
({\bf r})v^{{\bf q}\ast}_{\bf k}({\bf r})e^{2i{\bf q \cdot r}} , \label{gap2}
\ee
$E^{\bf q}_{\bf k}=(\xi_{\bf k+q}-\xi_{\bf q-k})/2 \pm \sqrt{(\xi_{\bf k+q}+
\xi_{\bf q-k})^2/4+
|\Delta^{\bf q}|^2} 
\equiv E_D \pm \sqrt{E_A^2+|\Delta^{\bf q}|^2}$,
where $E_D$ is the energy difference and $E_A$ the average energy. 
The $\pm$ holds for the particle and hole branch, respectively, and the particle branch eigenfunctions are
$|\tilde{u}^{\bf q}_{\bf k}|^2,|\tilde{v}^{\bf q}_{\bf k}|^2 = (1\pm E_A/\sqrt{E_A^2 + |\Delta^{\bf q}|^2})/2$.
The Hamiltonian transformed under the Bogoliubov 
transformation leading to (\ref{bdg}) has to be positive definite. This means
that one should use the solutions for which $E^{\bf +q}_{\bf k}>0$, i.e. $\min(\sqrt{E_A^2+|\Delta^{\bf q}|^2})
= |\Delta^{\bf q}|> E_D$. Remarkably,
this condition is the same as obtained using the Landau criterion.

In the BCS ansatz, a common momentum {\bf q} can be added to all particles, 
leading to correlations of the type
$\langle c_{{\bf k}+{\bf q}}^\dagger  c_{-{\bf k}+{\bf q}}^\dagger\rangle$. 
The superfluid net momentum becomes $2{\bf q}$. 
One can formally calculate
this obvious result also by using the plane wave ansatz
 $u_{\bf k} = |u_{\bf k}| e^{i({\bf k}+{\bf q})\cdot {\bf r}}$, 
$v_{\bf k} = |v_{\bf k}| e^{i({\bf k}+{\bf q})\cdot {\bf r}}$ \cite{BdG} 
(Eq.(\ref{gap2}) with $\phi = 1$) 
and introducing 
an (unnormalized) order parameter wave 
function $\Delta^{\bf q} ({\bf r}) = e^{i2{\bf q \cdot r}} C$, where $C$ is 
given by Eq.(\ref{gap2}) to be a constant in ${\bf r}$. Expectation values like momentum (${\bf p} =-i 
\partial/\partial {\bf r}$) can be calculated:
 $\langle {\bf p} \rangle= \langle \Delta^{\bf q}({\bf r}) |-i
\partial/\partial {\bf r}|
\Delta^{\bf q}({\bf r})\rangle/ \langle \Delta^{\bf q}({\bf r}) |\Delta^{\bf q} ({\bf r})\rangle = 2{\bf q}$. The order parameter wave function
is defined in the spirit of (but not with a one-to-one correspondence to) the Ginzburg-Landau theory with
a space 
dependent wave function whose absolute value equals the gap. In case of Fermionic atoms the
Ginzburg-Landau approach has been used to describe harmonic confinement \cite{GL} and vortices \cite{vortex}.
For the periodic potential we introduce the order parameter wave function in the form $\ket{\Delta^{\bf q}({\bf r})}
= \sum_{\bf k} \ket{\Delta^{\bf q}_{\bf k}({\bf r})}$, where using Eq.(\ref{gap2}),
\begin{equation}
\ket{\Delta^{\bf q}_{\bf k}({\bf r})}=F({\bf k,q}) \ket{\phi_{\bf k+q} e^{i 
({\bf k+q}) {\bf r}}}\ket{\phi^\dagger_{\bf k-q} e^{-i ({\bf k-q}) {\bf r}}}  \label{gapbk}
\end{equation}
and $F({\bf k,q}) = |g|[1-2f(E^{\bf +q}_{\bf k})]\tilde{u}^{\bf q}_{\bf k} 
\tilde{v}^{{\bf q}\ast}_{\bf k}$.
We calculate the superfluid velocity using 
$ 
\langle {{\bf v}_S} \rangle = {\cal N} \bra{\Delta^{\bf q}({\bf r})}\dot{\bf r}\ket{\Delta^{\bf q}({\bf r})},
$
where ${\cal N} = \langle{\Delta^{\bf q}({\bf r})}|{\Delta^{\bf q}({\bf r})}\rangle^{-1}$. 
Using
$\langle \dot{\bf r} \rangle_{\phi^\dagger_{\bf k-q} e^{-i({\bf k-q})\cdot{\bf r}}}
=-\frac{1}{\hbar} \frac{d}{d{\bf k}} \xi_{\bf k-q}$ 
and the tight-binding energy dispersion relation 
the superfluid velocity becomes
\begin{eqnarray}
\langle {v_{xS}}\rangle &=& {\cal N} \sum_{\bf k} |F({\bf k,q})|^2 
\frac{Ja}{\hbar} \cos{k_xa}\sin{qa} \label{vs} \\
&=& \frac{Ja}{\hbar} \sin{(qa)} 
{\cal N} \sum_{\bf k} \left| \frac{\left[ 1-2f(E^{\bf q}_{\bf k}) \right] 
\Delta^{\bf q}}{\sqrt{E_A^2+|\Delta^{\bf q}|^2} } \right|^2 
\cos{k_xa}.
\nonumber
\end{eqnarray}
The superfluid velocity for selected parameters is shown in 
Fig.\ \ref{landaucriteria}. We have also calculated the thermal quasiparticle contribution but is turns out
to be negligible.

In the composite boson limit, one could describe the center-of-mass movement of the composite particle 
by defining $J^* = J(m \rightarrow 2m)$. 
In order to give a simple estimate for the effect of the Fermi statistics, we interpret $|F({\bf k,q})|^2\sim|F({\bf k})|^2$ in Eq.(\ref{vs}) 
as reflecting the internal wavefunction of the pair in the composite boson limit, c.f.\ \cite{nozi,legget}. 
The average velocity for the bosons becomes 
$
\langle{v_{xB}}\rangle \propto \frac{J^* a}{\hbar} \sin{qa} \sum_{\bf k} |F({\bf k})|^2 \cos{k_xa}$. 
If the pairs were extremely strongly bound, the internal wave function in real space is a delta-function, 
corresponding to a constant in k-space. This means $\langle{v_B}\rangle=0$ since the cosine integration in Eq.(\ref{vs}) 
would extend to the whole k-space with equal weight, i.e.\ there are no empty states in the Brillouin zone 
as required for Bloch oscillations. For on-site pairs, we use $|F(r)|^2 \propto \exp(-r^2/l^2)$ leading to 
$|F(k)|^2 \propto {\cal N} \exp(-l^2k^2/4)$, therefore the suppression factor for the Bloch oscillations becomes $S \sim {\cal N} \int dk  \exp(-l^2k^2/4) \cos{ka}$, where $l$ is the pair size.
As a rough estimate for the average velocity we thus obtain
\be
\langle{v_{xB}}\rangle \sim S \frac{J^*a}{\hbar} \sin{(F_xta/\hbar)} . \label{blim}
\ee
This is shown in Fig.\ \ref{landaucriteria} for pair sizes $l=a/3$ and $l=a/4$. It gives 
an order-of-magnitude estimate, approaching the results given by the BCS algebra. 

Another way of treating the composite boson limit is to derive a Gross-Pitaevskii type of equation for the composite bosons with $M=2m$ and with a repulsive non-linear interaction term 
$n_B U_B= n_B 4\pi\hbar^2 a_B/M$, $a_B=2a_s$ where $a_s$ is the renormalized s-wave scattering length \cite{grossp}. 
If the non-linear term is small compared to the Bloch energy $E_B= h^2/(Ma^2)$, the nonlinearity leads only to a change in the band width $J$ 
\cite{bose1}. Therefore, composite bosons oscillate but with a modified amplitude. Large non-linearity would not allow Bloch oscillations, corresponding to a large suppression factor in the above discussion.  
Note that the Landau criterion for a superfluid Bose gas gives the critical 
velocity $v_{sound}=\sqrt{\frac{U_Bn_B}{M}}$ which is orders of 
magnitude bigger than Eq.(\ref{blim}) for half filling and parameters in 
Fig.\ \ref{landaucriteria}. Problems arise only in the extremely empty lattice limit.

In summary, we have defined a set of tools for qualitative and quantitative description of Bloch oscillations for the BCS-BEC crossover regime. The amplitude of the oscillations decreases when the crossover is scanned, in general due to the shrinking of the bandwidth. However, the change from the normal to the superfluid state description leads to a drastic change in the amplitude. This is due to smoothening of the Fermi-edge by pairing. Bloch oscillations could be used for exploring pairing correlations since any localization in space (pair size) leads to broadening in momentum which suppresses the amplitude in the same way as band filling in the non-interacting gas. Achievement of superfluidity is still a great challenge, but even at $T>>T_c$, the effect of collisions on Bloch oscillations can be studied producing information useful for applications of Bloch oscillations such as production of Terahertz radiation
\cite{thz,rusos}. Observation of oscillating fermionic atoms in optical lattices would contribute to the quest for a steadily driven fermionic Bloch oscillator. 
     
{\it Acknowledgements} We thank T.\ Esslinger and M.\ K\"ohl for useful discussions, and Academy of Finland (Project Nos.\ 53903, 48445) and ESF (BEC2000+ programme) for support.

\end{document}